# Concurrent Cyber Physical Systems: Tensor State Space Representation

Garimella Rama Murthy.
Associate Professor,
IIIT-Hyderabad, Gachibowli,
Hyderabad-500032

**Abstract**

In this research paper, state space representation of concurrent, linearly coupled dynamical systems is discussed. It is reasoned that the Tensor State Space Representation (TSSR) proposed in [Rama1] is directly applicable in such a problem. Also some discussion on linearly coupled, concurrent systems evolving on multiple time scales is included. Briefly new ideas related to distributed signal processing in cyber physical systems are included.

1. **Introduction:**
Electrical networks containing passive elements ( such as resistors, inductors, capacitors ) were investigated for transient as well as equilibrium behavior. Mathematical modeling of such networks was facilitated using ordinary, constant coefficient differential equations. Such an approach provided the input-output description of Linear Time Invariant (LTI) systems. Using Laplace transform, constant coefficient ordinary differential equations were converted into algebraic equations and the transfer function provided the complete description of LTI systems.
Kalman for the first time realized that state space description of linear time varying as well as LTI systems was very useful and convenient. With this innovative idea, modern control theory was developed extensively by many researchers using linear algebraic techniques. Concepts such as "controllability", "observability", "stability" were proposed using solution of state space equations [DoB].
In recent years, cyber physical systems were proposed as the integration of physical world and the cyber world ( i.e physical processes and computation/communication/control were integrated ). In cyber physical systems, taking care of "concurrency" is extremely important [Lee]. Mathematical modeling of cyber physical systems is being attempted by various researchers. This research paper is an attempt in that direction (i.e. mathematical modeling ). Specifically, this research paper proposes state space representation of concurrent, linear cyber physical systems.

This research paper is organized as follows. In Section 2, one dimensional linear concurrent systems are considered and their state space representation is discussed. In Section 3, multi-dimensional linear concurrent systems are considered and it is shown that the Tensor State Space Representation [TSSR], first proposed in [Rama1] is extremely useful. In Section 4, some interesting ideas related to distributed signal processing in cyber physical systems are proposed. In Section 5, state space representation of linear dynamical systems on different time scales is discussed. Essential idea of arriving at a global clock is proposed. The research paper concludes in Section 6.

2. **One Dimensional Linear Concurrent Systems: State Space Representation:**

Consider various dynamic phenomena which are concurrently evolving in time. Let the "state of each of them be represented by a single (scalar) function of time. Each of these functions can be considered to represent a physical process evolving continuously in time. For the purposes of concreteness, let us represent the physical processes as

$$\{ x_i(t): 1 \leq i \leq M \} \ldots\ldots\ldots\ldots(2.1)$$

These concurrent physical processes can be represented by means of an M-dimensional vector i.e.

$$X(t) = \begin{bmatrix} x_1(t) \\ x_2(t) \\ \vdots \\ x_{M-1}(t) \\ x_M(t) \end{bmatrix} \ldots\ldots\ldots\ldots(2.2)$$

- This vector can be considered to represent the 'state' of M-concurrent coupled scalar valued processes. This representation has the same advantages as that of state space representation of dynamical systems ( linear or non-linear ). Specifically one of the advantages is that this representation enables studying multiple input, multiple output systems, which may be linear or non-linear, time invariant or time varying.

Thus, most generally, we have the following state space representation of multiple concurrent systems ( that are coupled to one another ). Let U(t) represent the input to these coupled systems,

$$\frac{d}{dt} X(t) = f(X, U, t) \ldots\ldots\ldots\ldots(2.3)$$

$$Y(t) = g(X, U, t) \ldots\ldots\ldots\ldots(2.4)$$

- If the concurrent systems are linearly coupled to one another, we have the following state equations

$$\frac{d}{dt} X(t) = A(t)X(t) + B(t)\,\dot{U}(t) \ldots\ldots\ldots (2.5)$$

$$Y(t) = C(t)X(t) + D(t)\,U(t) \ldots\ldots (2.6)\,.$$

- In the case of linearly coupled concurrent time-invariant systems, we have the following state space representation:

$$\frac{d}{dt} X(t) = A\,X(t) + B\,\dot{U}(t) \ldots\ldots (2.7)$$

$$Y(t) = C\,X(t) + D\,U(t) \ldots\ldots\ldots (2.8)\,.$$

**Note**: In case where A is a diagonal matrix, the concurrent systems are decoupled. In the case where "state" of each of the concurrent processes is scalar valued, the linear coupling is "heterogeneous ( i.e coupling constants that are entries of matrices A,C depend on the process under consideration ).

**Note:** In the following discussion, we only consider linearly coupled systems.

- Now, we consider the case where each system is represented by not a scalar valued state but a vector valued state. We specifically consider the case where concurrent systems are linearly coupled.

- Let the state vectors $\{ X_i(t) : 1 \leq i \leq M \}$ be "augumented" to arrive at the M x M STATE MATRIX, $\bar{Z}(t)$ of coupled concurrent systems ( For convenience, we assume that the number of concurrent systems is same as the dimension of state vectors. This assumption can easily be relaxed ). Specifically, we have the STATE MATRIX,

$$\bar{Z}(t) = [X_1(t): X_2(t): \ldots : X_M(t)]\ldots\ldots\ldots(2.9)$$

Hence the state equations of such linearly coupled concurrent system are given by ( $\overline{W}(t)$ is the output matrix ).

$$\frac{d}{dt}\bar{Z}(t) = A(t)\bar{Z}(t) + B(t)\,\dot{\bar{U}}(t) \ldots\ldots (2.10)$$

$$\overline{W}(t) = C(t)\bar{Z}(t) + D(t)\,\bar{U}(t) \ldots\ldots (2.11)$$

Naturally, we also have the following state space representation of linear, time-invariant concurrent systems.

$$\frac{d}{dt}\bar{Z}(t) = A\,\bar{Z}(t) + B\,\dot{\bar{U}}(t) \ldots\ldots (2.12)$$

$$\overline{W}(t) = C\ \bar{Z}(t) + D\ \overline{U}(t) \ldots\ldots (2.13)$$

**Note:** In this case, the state of each of the concurrent processes is vector valued ( i.e. they are the columns of the matrix $\bar{Z}(t)$ ). The linear coupling matrices ( corresponding to concurrent processes ) are "homogeneous" ( i.e. the coupling matrices A,C are the same for all the state vectors that are columns of $\bar{Z}(t)$ ).

- Now, we consider linear, concurrent coupled systems evolving in discrete time ( with the state of each system being vector valued ). The state space representation is provided below.

$$\bar{Z}(n+1) = A(n)\dot{\bar{Z}}(n) + B(n)\overline{U}(n) \ldots (2.14)$$
$$\overline{W}(n) = C(n)\bar{Z}(n) + D(n)\overline{U}(n) \ldots (2.15)$$

The state space representation, when the system is time invariant can easily be derived from the above one.

- It is most natural to represent "heterogeneously" coupled linear concurrent systems using an appropriate representation. Thus, in this case, the most natural thing to do is to consider 3-d arrays as the coupling arrays instead of the coupling matrices $\{A, B, C, D\}$. For instance let us consider the case where the array A is three dimensional and state is a matrix. In this case, we first compute the outer product i.e.

$$A_{i_1,i_2,i_3} \cdot X_{j_1,j_2} = E_{k_1,k_2,k_3,k_4,k_5}.$$

Formally, in the above equation, we considered the outer product between the tensors "A" and "X". Using appropriate "contraction" operation over the suitable indices, the inner product between them is determined ( i.e. five dimensional array E is reduced to a two dimensional array or a matrix ). The contraction depends on the heterogeneous coupling that is required.

In the most general case, the arrays $\{A, B, C, D\}$ are tensors of desired order and degree.

- Now, we naturally consider "heterogeneously coupled" linear concurrent systems whose state is a matrix / 3-d array / multi-dimensional array i.e. a tensor. The challenging problem associated with such systems is the STATE SPACE REPRESENTATION ( with AUGUMENTED state / input tensors of linear concurrent systems ). We address this representation problem in the following section.

## 3. Multi-Dimensional Linear Concurrent Systems: Tensor State Space Representation:

In his research monograph [Rama1], the author discusses the concept of Tensor State Space Representation (TSSR) of certain multi-dimensional systems. We now briefly explain Tensor State Space Representation in the following discussion. We first consider systems evolving in discrete time. The discussion can be naturally extrapolated for systems evolving in continuous time.

The main idea behind TSSR is to replace first order / second order tensors ( i.e. vector, matrix ) in the one dimensional state space representation i.e.

$$X(n+1) = A(n)\dot{X}(n) + B(n)U(n)$$

$$Y(n) = C(n)X(n) + D(n)U(n)$$

by higher order tensors. Thus, we have the following Tensor State Space Representation (TSSR) of certain multi-dimensional linear systems:

$$X_{(i_1,\ldots,i_r)}(n+1) = A_{(i_1,\ldots,i_r\,;\,j_1,\ldots,j_r)}(n)\,X_{(j_1,\ldots,j_r)}(n) + B_{(i_1,\ldots,i_r\,;\,j_1,\ldots,j_p)}(n)\,U_{(j_1,\ldots,j_p)}(n)$$

$$Y_{l_1,\ldots,l_s}(n) = C_{l_1,\ldots,l_s;\,j_1,\ldots,j_r}(n)\,X_{(j_1,\ldots,j_r)}(n) + D_{(l_1,\ldots,l_s\,;\,j_1,\ldots,j_p)}(n)\,U_{(j_1,\ldots,j_p)}(n)$$

Where **A**(n) is an m-dimensional tensor of order 2r ( called the state coupling tensor), **X**(n) is the state of dynamical system at the discrete time index 'n', whereas **X**(n+1) is the state of the system at the discrete time index 'n+1'. Furthermore **B**(n) is an m-dimensional tensor of order 'r+p' ( called the input coupling tensor ), **Y**(n) is the output tensor of dimension 'm' and order 's'. **U**(n) is an m-dimensional input tensor ( varying with discrete time index) of order 'p' and **C**(n) ( called state coupling tensor to the output dynamics ) is an m-dimensional tensor of order (s+r), **D**(n) is the input coupling tensor to the output dynamics of dimension 'm' and order 's+p'.

**Remark 1:** In the state space representation of one dimensional systems given in equations (2.5), (2.6), by replacing the one/two dimensional tensors with suitable higher order tensors, we arrive at the Tensor State Space Representation of linear, concurrent, "heterogeneously" coupled continuous time systems. Detailed equations are avoided for brevity.

**Remark 2:** So far, we were successful in arriving at the state space representation of "heterogeneously coupled", linear concurrent systems ( by augumenting the state tensors ). This representation is extremely useful because the results ( such as solution of state equations, concepts such as "controllability", "observability", "stability" etc ) available for one dimensional systems can be naturally generalized to "heterogeneously" coupled, linear concurrent systems ( whose state tensors are "augumented" ). Thus design and analysis of such systems is facilitated using the tensor state space representation.

### 4. Distributed Signal Processing : Cyber Physical Systems:

In the recent years, many applications motivated the design of distributed signal processing algorithms ( for instance smart grid design ). There were two main efforts:

- Centralized computation of say a quantity like "mean".
- Decentralized computation of say "mean" ( using the "consensus" algorithm ).

- The author realized ( during the invited talk of Jose Moura [Mou] ) that in many cyber physical systems ( such as wireless sensor networks ) the computation of interesting quantity ( say "mean" values of the associated natural / artificial phenomenon ) is "partly distributed" and "partly centralized".

Thus, the algorithms ( such as statistical estimation techniques using the concepts such as "innovations" process ) developed for purely centralized / purely decentralized case should be modified for the case that naturally arises in the design of many cyber physical systems [Rama2].

### 5. Linear Dynamical Systems on Different Time Scales:

In natural as well as artificial systems, the physical processes are evolving on multiple time scales. In most cases they interact with each other. Thus, we have coupled/interacting distributed linear dynamical systems on different time scales.

- Goal: To represent such coupled, concurrent linear dynamical systems on multiple time scales and solve for the state trajectories.

Consider two scalar valued ( state ), linearly coupled concurrent systems evolving on different time scales. Let the dynamics be represented by the following equation:

$$\begin{bmatrix} x_1(n) \\ x_2(n) \end{bmatrix} = \begin{bmatrix} a_{11} & a_{12} \\ a_{21} & a_{22} \end{bmatrix} \begin{bmatrix} x_1\left(\frac{n}{c_1}\right) \\ x_2\left(\frac{n}{c_2}\right) \end{bmatrix} + \begin{bmatrix} b_{11} & b_{12} \\ b_{21} & b_{22} \end{bmatrix} \begin{bmatrix} u_1\left(\frac{n}{c_1}\right) \\ u_2\left(\frac{n}{c_2}\right) \end{bmatrix},$$

where $\{ c_1, c_2, d_1, d_2 \}$ are positive integers larger than one ( i.e. These scalar valued discrete time systems are evolving on different time scales are coupled ). Now, let the Least Common Multiple (L.C.M) of $\{ c_1, c_2 \}$ be 'd' i.e.

L.C.M $\{ c_1, c_2 \} = d$ $and$ $n' = \frac{n}{d}$.

Thus, we have that

$$\begin{bmatrix} x_1(dn') \\ x_2(dn') \end{bmatrix} = \begin{bmatrix} a_{11} & a_{12} \\ a_{21} & a_{22} \end{bmatrix} \begin{bmatrix} x_1(f_1 n') \\ x_2(f_2 n') \end{bmatrix} + \begin{bmatrix} b_{11} & b_{12} \\ b_{21} & b_{22} \end{bmatrix} \begin{bmatrix} u_1(f_1 n') \\ u_2(f_2 n') \end{bmatrix},$$

Now using L.C.M $\{ d, f_1, f_2 \}$, a common global clock is defined to describe the system on a single time scale. Using standard techniques, the state equations on a global clock are solved.

**Remark 3:** The generalization of above idea to the case where the state of coupled, concurrent systems are tensor valued is straight forward. Details are avoided for brevity

- More interestingly, results of multi-rate signal processing are capitalized to derive new results for multiple, linearly coupled concurrent systems evolving on multiple time scales.

## 6. Conclusions:

In this research paper, state space representation of linearly Coupled concurrent dynamical systems is discussed. It is shown that Tensor State Space Representation (TSSR) proposed in [Rama1] is very useful in such an effort. Also state space representation of linear concurrent dynamical systems evolving on different time scales is briefly discussed.